\documentclass[preprint,trackchanges]{aastex62}
\hypersetup{linkcolor=blue,citecolor=blue,filecolor=cyan,urlcolor=blue}
\shorttitle{Recurrent Two-Sided Loop Jets}
\shortauthors{Yang et al.}
\begin{document}

\title{ Recurrent Two-Sided Loop Jets Caused by Magnetic Reconnection between
Erupting Minifilaments and Nearby Large Filament}

\correspondingauthor{Bo Yang}
\email{boyang@ynao.ac.cn}

\author{Bo Yang}
\affil{ Yunnan Observatories, Chinese Academy of Sciences, 396 Yangfangwang, Guandu District, Kunming, 650216, People's Republic of China}
\affiliation{Center for Astronomical Mega-Science, Chinese Academy of Sciences, 20A Datun Road, Chaoyang District, Beijing, 100012, People's Republic of China}
\affiliation {Key Laboratory of Solar Activity, National Astronomical Observatories of Chinese Academy of Science, Beijing, 100012, People's Republic of China }

\author{Jiayan Yang}
\affil{ Yunnan Observatories, Chinese Academy of Sciences, 396 Yangfangwang, Guandu District, Kunming, 650216, People's Republic of China}
\affiliation{Center for Astronomical Mega-Science, Chinese Academy of Sciences, 20A Datun Road, Chaoyang District, Beijing, 100012, People's Republic of China}

\author{Yi Bi}
\affil{ Yunnan Observatories, Chinese Academy of Sciences, 396 Yangfangwang, Guandu District, Kunming, 650216, People's Republic of China}
\affiliation{Center for Astronomical Mega-Science, Chinese Academy of Sciences, 20A Datun Road, Chaoyang District, Beijing, 100012, People's Republic of China}
\affiliation {Key Laboratory of Solar Activity, National Astronomical Observatories of Chinese Academy of Science, Beijing, 100012, People's Republic of China }

\author{Zhe Xu}
\affil{Key Laboratory of Dark Matter and Space Astronomy, Purple Mountain Observatory, Chinese Academy of Sciences, Nanjing 210034, People's Republic of China }

\author{Junchao Hong}
\affil{ Yunnan Observatories, Chinese Academy of Sciences, 396 Yangfangwang, Guandu District, Kunming, 650216, People's Republic of China}
\affiliation{Center for Astronomical Mega-Science, Chinese Academy of Sciences, 20A Datun Road, Chaoyang District, Beijing, 100012, People's Republic of China}

\author{Haidong Li}
\affil{ Yunnan Observatories, Chinese Academy of Sciences, 396 Yangfangwang, Guandu District, Kunming, 650216, People's Republic of China}
\affiliation{Center for Astronomical Mega-Science, Chinese Academy of Sciences, 20A Datun Road, Chaoyang District, Beijing, 100012, People's Republic of China}

\author{Hechao Chen}
\affil{ Yunnan Observatories, Chinese Academy of Sciences, 396 Yangfangwang, Guandu District, Kunming, 650216, People's Republic of China}
\affiliation{Center for Astronomical Mega-Science, Chinese Academy of Sciences, 20A Datun Road, Chaoyang District, Beijing, 100012, People's Republic of China}
\affiliation{University of Chinese Academy of Sciences, 19A Yuquan Road, Shijingshan District, Beijing, 100049, People's Republic of China}



\begin{abstract}
Using high spatial and temporal data from the New Vacuum Solar Telescope (NVST) and the \emph{Solar Dynamics Observatory} (\emph{SDO}),
we present unambiguous observations of recurrent two-sided loop jets
caused by magnetic reconnection between erupting minifilaments and nearby large filament.
The observations demonstrate that three two-sided loop jets, which ejected along the large filament in opposite directions,
had similar appearance and originated from the same region.
We find that a minifilament erupted and drove the first jet. It reformed at the same neutral line later, and then underwent partial and total eruptions,
drove the second and third jets, respectively. In the course of the jets, cool plasma was injected into the large filament.
Furthermore, persistent magnetic flux cancelation occurred
at the neutral line under the minifilament before its eruption and continued until the end of the observation.
We infer that magnetic flux cancellation may account for building and then triggering the minifilament
to erupt to produce the two-sided loop jets. This observation not only indicates that two-sided loop jets can be driven by minifilament eruptions,
but also sheds new light on our understanding of the recurrent mechanism of two-sided loop jets.

\end{abstract}


\section{Introduction} \label{sec:intro}
It is generally accepted that omnipresent jet activities of various spatial and temporal scales in the solar atmosphere are caused by
magnetic reconnection between emerging magnetic flux and preexisting ambient coronal magnetic fields \citep{shibata92,shibata94,shimojo96,jiang07,par10,yang11,li12,rao16}.
\citet{shibata94} suggested that when a bipolar emerging flux emerges from below the solar surface into the corona
and interacts with a vertical or oblique coronal field, an anemone jet will be initiated; and when a bipolar
emerging flux intrudes into and continuously interacts with a horizontal coronal field, a two-sided loop jet will be initiated.
On the basis of magnetic-reconnection jet model, both anemone jet and two-sided loop jet
were successfully modeled by the magnetohydrodynamic (MHD) simulations \citep{yok95,yok96}.

Typical jets are single field-directed jets that consist of a bright base and a single spire, along which heated plasmas are ejected outward.
\citet{shibata94} called these jets as anemone jets.
Based on morphological descriptions of coronal jets, \citet{moore10} introduced the concept of ``standard jets"
and ``blowout jets" to describe the single spire jets. One difference is that the spire stays narrow in what they call standard jets,
while it is observed to become broad and multi-stranded in what they call blowout jets. They assumed that both standard and blowout jets
resulted from emerging flux and the bright point frequently observed on the side of the base of jets came from the
reconnection between the emerging/emerged bipole and the ambient field. This is similar to what was assumed by \citet{shibata92} and \citet{yok95}.
In the model of \citet{moore10}, for standard jets, the idea was that this reconnection made a narrow jet spire, and then the reconnection stopped.
For blowout jets, the idea was that this initial reconnection resulted in the emerging/emerged bipole erupting outward, making a broad spire.
In addition, they found that the blowout eruption of the bipole field and the surrounding field often carrying a filament of cool plasma.
Subsequently, more and more observations confirmed that many blowout jets are driven by minifilament eruptions
\citep{hong11,hong16,hong17,chen12,shen11,shen12,shen17,yang12,moore13,adams14,sterling16,li17}.
In particular, \citet{sterling15} investigated 20 near-limb polar coronal hole jets and  morphologically
classified 14 as blowout, 5 as standard, and 1 as ambiguous. They found that all the jets originated
from minifilament eruptions. On the basis of their observations, they suggested that standard jets and blowout jets are fundamentally
the same phenomenon, and the eruption's strength and whether the erupting filament escapes the base determine
whether the jet has a standard or a blowout morphology. The blowout jets model of \citet{moore10}
would explain the filament eruption, but not the location of the initial bright point. However, the observation of \citet{sterling15}
argued that the bright point was from the reconnection below the erupting minifilament
(just like normal flares occurring below erupting large filaments)
rather than by reconnection with the external ambient field.

Two-sided loop jets are bi-directional jets or transient loop brightenings that consist of two roughly anti-parallel
spires developing from a central bright region. Currently, only a few observations of two-sided loop jets, which support the
emerging reconnection scenario \citep{yok95,yok96}, have been reported \citep{kun98,kun99,alex99,jiang13,zheng18}.
In contrast to the emerging reconnection scenario,
\citet{tian17} reported that the observed two-sided loop jets are caused by magnetic reconnection between two adjacent filamentary threads.
\citet{yang16} found that the reconnection between two adjacent filamentary threads formed not only a bi-directional jet but also a filament.
More recently, \citet{sterling19} presented, for the first time, that a two-sided loop jet was resulted from the reconnection between
an erupting minifilament field and the overlying nearly horizontal loop field. Soon after, \citet{shen19} reported a two-sided loop jet that was driven by
the interaction of an erupting minifilament field and its overlying large filament field.
Similar to the case of anemone jets, it seems that minifilament eruptions should also play
an important role in the formation of two-sided loop jets. Therefore, more observations of minifilament
eruptions driving two-sided loop jets need to be further investigated. In this study,
with high-quality data obtained by the New Vacuum Solar Telescope \citep[NVST;][]{liu14} and \emph{Solar Dynamics Observatory}
\citep[\emph{SDO};][]{pes12}, we present unprecedented details on the initiation and evolution
of recurrent two-sided loop jets caused by magnetic reconnection between
erupting minifilaments and nearby large filament. This observation sheds new
light on our understanding of the physical nature and the recurrent mechanism of two-sided loop jets.

\section{Observations and Data Analysis}
The recurrent two-sided loop jets occurred in NOAA active region (AR) 11692 on 2013 March 14
and were well-observed by the Atmospheric Imaging Assembly \citep[AIA;][]{lem12}
and the Helioseismic and Magnetic Imager \citep[HMI;][]{sch12} on board the \emph{SDO}. This event was also observed
by the Global Oscillation Network Group (GONG) at the National Solar Observatory (NSO).
AIA takes the full-disk images of the Sun in seven extreme ultraviolet(EUV)
wavelengths at 12 s cadence with 0.$\arcsec$6 pixel$^{-1}$. Here, we mainly used the 304 \AA\ (\ion{He}{2}, 0.05 MK)
, 171 \AA\ (Fe {\sc ix}, 0.6 MK ), and 94 \AA\ (\ion{Fe}{18}, 7 MK) images to study this event.
HMI measures the full-disk line-of-sight (LOS) magnetic field for the \ion{Fe}{1} absorption line at 6173 \AA\,,
with a cadence of 45 s and a spatial resolution of 1.$\arcsec$0. GONG provides full-disk $H_{\alpha}$
line center images with a pixel size of 1.$\arcsec$05 and a cadence of 1 minute.
All images, taken from the AIA, HMI, and GONG were then aligned by differentially rotating to the reference time of
03:00 UT on 2013 March 14.

On 2013 March 14, AR 11692 was also observed by the NVST in $H_{\alpha}$ 6562.8 \AA\ and TiO from 00:59:00 UT to 05:30:00 UT.
The $H_{\alpha}$ and the TiO images have a spatial resolution of 0.$\arcsec$33 and 0.$\arcsec$07, respectively.
The time cadence of these images is 12 s.
By subtracting dark current, correcting with flat field, and carrying out speckle masking,
the raw data were reconstructed to high resolution images \citep{xiang16}.
All of the high resolution images were co-aligned with each other by implementing
a high accuracy solar image registration procedure \citep{yang15}. Then, the \emph{SDO} and NVST images
were co-aligned with each other by carrying out an automatic mapping approach developed by \citet{ji19}.
Furthermore, the \emph{Reuven Ramaty High Energy Solar Spectroscopic Imager} \citep[\emph{RHESSI};][]{lin02}
were also utilized. The CLEAN method \citep{hur02} was used
to reconstruct the \emph{RHESSI} image at the energy bands of 6-12 keV,
with an integration time of 3 minutes (02:56-02:59 UT).

\section{Results}
\subsection{The Large Filament and the Minifilament}
Figure 1 shows snapshot images of the concerned large filament, ``LF", and the minifilament, ``MF" at 02:40 UT on 2013 March 14.
LF appears as an inverse S-shaped segment and has an apparent length of about 200 Mm (Figure 1({\it a}\sbond{\it c})).
Due to the limitation of the field-of-view (FOV), LF is not completely observed by NVST.
Fortunately, the fine structure of the main body of LF is still captured by the NVST
high-resolution $H_{\alpha}$ observations. It is clear from the NVST high-resolution $H_{\alpha}$ image (Figure 1({\it c}))
that LF consists of numerous long thin threads, which may represent separate flux tubes \citep{eng98}.
When contours of HMI magnetograms were superimposed on the simultaneous 171 \AA\ image (Figure 1({\it b})), it
became clear that the northeast end of LF roots at a positive-polarity region, while its southwest end roots at a negative-polarity region.
In particular, there are some conspicuous left-skewed coronal loops overlying LF and connecting opposite polarities on its two sides. These observational features
indicate that LF is dextral \citep{mar98}. To the left side of LF, MF appears as a fibril-like structure and has an apparent length of about 15 Mm.
From magnetogram overlaid onto the NVST high-resolution $H_{\alpha}$ image (Figure 1({\it d})), it is apparent that MF
roughly separates opposite-polarity magnetic fields with the northeast end rooted in a positive-polarity region
and the southwest end rooted in a negative-polarity region.
When viewed from the positive-polarity side of MF, the axial field
component points to the right. Hence, MF is identified as dextral \citep{mar98},
which is consistent with LF.

\subsection{The First Two-Sided Loop Jet}
The eruption of MF and the formation of the first two-sided loop jet,``J1", are shown in Figure 2.
Before J1 onset, at about 02:55 UT a brightening (as indicated by the blue arrows) appeared
beneath MF as it started to rise (Figure 2({\it a}) and Figure 2({\it d})).
Subsequently, it is clear from the NVST observations that the rising MF erupted toward
and merged into LF (Figure 2({\it c})). Meanwhile, the brightening became much more pronounced (Figure 2({\it b}))
and cool plasma  was ejected into LF (Figure 2({\it c})).
By scrutinizing the AIA 171 \AA\ observations (Figure 2({\it e})), it is found that remarkable bidirectional plasma ejections likely stemmed from the sites
where the rising MF (as denoted by the red arrow) collided with LF. The bidirectional plasma ejections further extended and ejected along LF in opposite directions
(Figure 2({\it f})\sbond({\it g})). Moreover, footpoint brightenings appeared at the positive end of LF (Figure 2({\it c}) and Figure 2({\it g})).
From the AIA 94 \AA\ observations, it is notable that the collision sites are characterized by an enhancement in brightness
and the bidirectional plasma ejections  are more like loop brightenings, which consist of two roughly anti-parallel
spires developing from the collision sites and extending along LF in opposite directions (Figure 2({\it h})\sbond({\it i})).
Thus, J1 is a typical two-sided loop jet. We also found that a \emph{RHESSI} HXR source
in the energy range of 6-12 keV appeared almost coincident with the strongest EUV brightenings at the collision sites (Figure 2({\it i})).
Along slice ``AB" marked in Figure 4(d), spacetime plots were constructed from
AIA 304 and 171 \AA\ images, respectively, and the result was provided in Figure 5({\it a})\sbond({\it b}).
The two components of J1 are clearly shown on the spacetime plot. Via performing linear fittings to the outer edge of the jet,
we found that the plasma ejected to the positive end of LF at a projected velocity of about 149 km s$^{-1}$ and to the
negative end of LF at about 166 km s$^{-1}$. Furthermore, the associated movie reveals that
the plasma ejected toward the positive end of LF partially returns to the negative end of MF (see also the spacetime plot)
and the plasma ejected toward the negative end of LF partially returns to the positive end of MF.
These observations strongly suggest that magnetic reconnection took place between the erupting MF
and its nearby LF, resulting in the birth and growth of J1.

\subsection{The Recurred Two-Sided Loop Jets}
After MF erupted, we found that successive two-sided loop jets recurred near 05:31 and 05:42 UT, respectively.
Figure 3 presents the birth and growth of the successive recurred two-sided loop jets,``J2" and ``J3".
After J1 eruption, the erupted MF slowly reformed at the same neutral line
at about 04:40 UT (Figure 3({\it a}) and Figure 3({\it d})).
The reformed MF shows a similar appearance and size with the erupted one. As MF reformed,
the NVST $H_{\alpha}$ movie shows that its southern part of the reformed MF gradually became thicker and darker.
Then after, this part of the reformed MF gradually lifted off, moved toward, and interacted with LF (Figure 3({\it b})\sbond({\it c})).
However, the main structure of the reformed MF still existed (as denoted by the arrow in Figure 3({\it c})).
Similar to previous MF eruption, significant brightenings (Figure 3({\it e})) appeared beneath the reformed MF as its southern part started to lift off.
However, these brightenings can only be seen on the AIA observations. It is remarkable from the AIA observations that bidirectional plasma ejections,
which also likely stemmed from the collided sites, appeared and ejected along LF in opposite directions
as soon as the rising part of the reformed MF collided with LF (Figure 3({\it f})). These observations imply that J2 is triggered
by the partial eruption of the reformed MF.

Immediately after the ejection of J2, J3 was initiated and further developed (Figure 3({\it h})\sbond({\it i})).
Unfortunately, the NVST high-resolution observations do not cover the birth and growth of J3.
However, the AIA observations captured the key information that could reveal the birth and growth of J3.
Prior to J3 initiated, at about 05:39 UT when the remaining MF started to rise, brightenings
(as denoted by the white arrow in Figure 3({\it g})) also appeared beneath it (as indicated by the yellow arrow).
Afterwards, the remaining MF interacted and merged with LF (see the associated movie). Likewise, bidirectional plasma ejections
stemmed from the collision sites and ejected along LF in opposite directions (Figure 3({\it h})\sbond({\it i})).
Finally, the associated AIA movie shows that the reformed MF completely disappeared.
The rising of the reformed MF and its merging with LF were also captured by the
GONG $H_{\alpha}$ observations (Figure 4({\it a})\sbond({\it c})).
It is evident from the GONG $H_{\alpha}$ observations that cool plasma was ejected into LF (Figure 4({\it c}))
as soon as the reformed MF erupted and merged into LF. Because of the low spatiotemporal resolution,
the GONG $H_{\alpha}$ observations could not distinguish between the partial and complete eruptions of the reformed MF.
After the total eruption of the reformed MF, the AIA 304 \AA\ observations (Figure 4({\it d})\sbond({\it f}))
reveal that partial magnetic field of LF changed the connectivity and reconnected to the negative end of the reformed MF.
Along slice ``CD" marked in Figure 4(d), spacetime plots were also
constructed from AIA 304 and 171 \AA\ images, respectively, and the result was provided in Figure 5({\it c})\sbond({\it d}).
J2 and J3 are clearly shown on the spacetime plots. For J2, the plasma ejected to the positive end of LF at a projected velocity of about 186 km s$^{-1}$
and to the negative end of LF at about 139 km s$^{-1}$; for J3, the plasma ejected to the positive end of LF at a projected velocity of about 145 km s$^{-1}$
and to the negative end of LF at about 125 km s$^{-1}$.
These observations indicate that magnetic reconnection occurred between the rising MF and LF, and
J3 should be triggered by the complete eruption of the remaining MF. Based on the observation, we suggest that
partial and then total eruptions of a single minifilament can result in the recurrence of two-sided loop jets.

\subsection{Photospheric Magnetic Field Evolution}
As described above, the initial eruption, the reformation, the subsequent partial and complete eruptions of MF clearly took place on the same magnetic neutral line (Figure 1({\it d})).
To explore the physical mechanisms of MF eruption and its reformation, we investigate the evolution of the photospheric magnetic field below MF.
Through checking the HMI data (Figures 6({\it a})\sbond({\it d})), we found that the area of the positive and negative magnetic flux patches,
which resided on either side of the neutral line, continuously decreased.
This means that significant flux cancellation occurred at the neutral line under MF.
It is notable from the NVST high-resolution TiO image (Figure 6({\it f})) that the canceling negative flux patch corresponds to a small pore.
The temporal evolution of the unsigned negative magnetic flux in the cancellation area is shown by the red curve in Figure 6({\it e}).
Here, we mainly measured the changes in the negative flux,
because it is more isolated and its evolution is easy to follow and measure. One can see from Figure 6({\it e}) that the unsigned negative flux persistently
decreased from about 02:00 UT and continued until the end of the observation. During this time interval, MF erupted, reformed, and then underwent partial and total eruptions.
This may confirm that both the eruption and reformation of MF occurred in conjunction with flux cancellation. It seems that the
continuing flux cancellation at the neutral line firstly triggered MF to erupt to drive J1; then,it gradually built up the field
holding MF's material, resulting in the reformation of MF; finally, it triggered the partial and complete eruption of the reformed MF,
generating J2 and J3, respectively. Magnetic flux cancelation, which  plays a key role in building up
and destabilizing the filament system, is frequently observed in the courses of filament formation and eruption \citep{van89,pan16,pan17,yan16,yang16,yang19,sterling18,sterling19}.
In particular, \citet{pan16,pan17} and \citet{sterling18,sterling19} suggested that magnetic flux cancellation is the source and trigger of the pre-jet minifilaments.
Our observations support the ideas of \citet{pan16,pan17} and \citet{sterling18,sterling19} that flux cancelation built up the minifilament system and triggered
its partial and complete eruptions, and then drove the production of the recurrent two-sided loop jets.

\section{Conclusion and Discussion}
In this paper, we study in detail the initiation and evolution of recurrent two-sided loop jets
caused by magnetic reconnection between erupting MF and nearby LF.
In our observations, we find that three two-sided loop jets originated from the same region
and had a closer resemblance to each other in the AIA observations.
The first two-sided loop jet was triggered by the interaction of a rising MF with nearby LF's magnetic field.
After MF erupted, a new MF was gradually reformed at the eruption source region.
The reformed MF firstly underwent a partial eruption. Its southern part lifted off and moved toward LF.
As a result, magnetic reconnection occurred between them and triggered the second two-sided loop jet.
Subsequently, the remaining MF underwent a complete eruption. Likewise, the erupted MF reconnected with LF's magnetic field
and triggered the third two-sided loop jet. Cool plasma was ejected into LF simultaneously with these jets.
The HMI observation reveals that persistent flux cancelation
occurred underlying MF before its eruption and lasted till the end of our observation.
The flux cancelation may firstly destabilize MF's field, leading to its eruption, then build up MF's field,
resulting in its reformation, and finally destabilize the reformed MF's field, causing its partial and complete eruptions.
Our observations shed new light on the understanding of the physical nature and the recurrent mechanism of two-sided loop jets.

This event provides solid observational evidence that recurrent two-sided loop jets can be driven by minifilament eruptions,
which is in line with previous observations that minifilament eruptions result in single-spire jets \citep{sterling15,sterling16,rao16}.
This is inconsistent with previous classical two-sided loop jet model \citep{yok95,yok96} and observations \citep{alex99,jiang13,zheng18}, which thought
two-sided loop jet is driven by magnetic reconnection between emerging bipoles and their overlying horizontal magnetic fields.
To date, only two observations of two-sided loop jet driven by minifilament eruptions have been reported.
The first observation was reported by \citet{sterling19} and then the second was reported by \citet{shen19}.
They also proposed minifilament-eruption models which could drive two-sided loop jets based on their observations.
Their models are very similar to the minifilament-eruption model that drives single-spire jets \citep{sterling15,sterling16}.
The key difference between them is that the single-spire jet model implies that the erupting minifilament system reconnecting with the ambient open field lines
can result in coronal mass ejections (CME) \citep{moore10,hong11,shen12,miao18,duan19}, while the two-sided loop jet model indicates that the erupting minifilament system
reconnecting with the overlying horizontal magnetic fields confines the eruptions of the jet plasma and the minifilament.
Therefore, no CME can be produced. In our observation, MF is not below LF
but on the left side of LF. However, the driving of the two-sided loop jet can fully be explained by
the models of \citet{sterling19} and \citet{shen19}. Our observation implies that no matter where the minifilament is located in,
two-sided loop jets would be generated as long as the minifilament erupts toward and reconnects with its nearby horizontal magnetic fields.

It is worth noting that both \citet{sterling19} and \citet{shen19} reported only a single two-sided loop jet.
However, previous observations \citep{jiang13,tian17,zheng18} and our observation here indicated that two-sided loop jet
often has a tendency to recur from the same neutral line. The jet models of \citet{sterling19} and \citet{shen19}
could not account for recurrent two-sided loop jets. The observational results presented in this paper provide
new insights into the mechanism of recurrent two-sided loop jets. Previous studies have shown that the continuous emergence of magnetic flux brought in successively,
the emergence of coronal loops, which continuously interact and reconnect with the overlying horizontal magnetic field,
resulting in the successive two-sided loop jets \citep{yok95,yok96,jiang13,tian17,zheng18}.
In our observation, MF erupted completely to form the first two-sided loop jet,
and then a new MF reformed at the same neutral line, a portion of the reformed MF erupted to produce the second two-sided loop jet,
and finally the remaining portion of the reformed MF erupted sometime later to form the third two-sided loop jet.
This process is always accompanied by magnetic flux cancellation.
This is consistent with the observations of \citet{pan16,pan17} that partial and total eruptions of a single minifilament
or the repeated formation and then eruption of the minifilament can drive recurrent single-spire jets.
Meanwhile, magnetic flux cancellation results in both the formation of the minifilament and its partial and total eruptions.

How the cool dense plasma is formed in the filament is still a controversial question in solar physics. Previous studies
have proposed several promising models accounting for the formation of cool dense plasma in the filament, such as the injection model \citep{wang99},
the levitation model \citep{rus94}, the evaporation-condensation model \citep{ant99,xia16}, and the reconnection-condensation model \citep{kan17}.
\citet{wang99} proposed that through a series of magnetic reconnection processes associated with flux cancellation at the photospheric neutral line,
chromospheric plasma can be injected into the filament channel in the form of jets or upflows \citep{wang99,liu05,wang18}. The present
observations also demonstrate that cool plasma was injected into the large filament in the courses of the two-sided loop jets.
Here, we provide a new way to understand how the flux cancellation supplies cool dense plasma
to the large filament. The magnetic structure of MF may {be built up} by the magnetic reconnection associated with flux cancellation
at the photospheric neutral line \citep{van89,yang16}. Meanwhile, cool dense plasma from the photosphere or chromosphere may be
lifted or injected into the magnetic structure of MF \citep{van89,rus94}. As the flux cancellation continues, MF is
destabilized, partial or total eruptions occur. The erupted MF further reconnects with the large filament. As a result, the cool dense plasma
carried by MF is injected into the large filament in the form of two-sided loop jets. Repeating this process,
cool dense plasma could be continually transported to the large filament.
\citet{mar98} suggested that magnetic flux cancellation is abundant under the large filament channel near the photospheric neutral line.
It seems that minifilament might be commonly formed and erupted at these locations. Erupting minifilament interacted with large filament
and injected cool plasma into it should be frequently occurred. However, more high resolution observations, such as the NVST, are needed
for further investigate.

\acknowledgments
We thank the referee for valuable suggestions and comments that greatly improved the paper.
The authors thank the NVST, \emph{SDO}, GONG, and \emph{RHESSI} science teams for providing the excellent data.
This work is supported by the Natural Science Foundation of China, under grants 11703084, 11633008, 11873088, and 11933009;
the CAS ``Light of West China" Program; the Open Research Program of the Key Laboratory of Solar Activity of Chinese Academy of Sciences (KLSA201809);
the CAS grant ``QYZDJ-SSW-SLH012"; and the grant associated with the Project of the Group for Innovation of Yunnan Province.

\newpage
\begin{figure}
\epsscale{1.}
\plotone{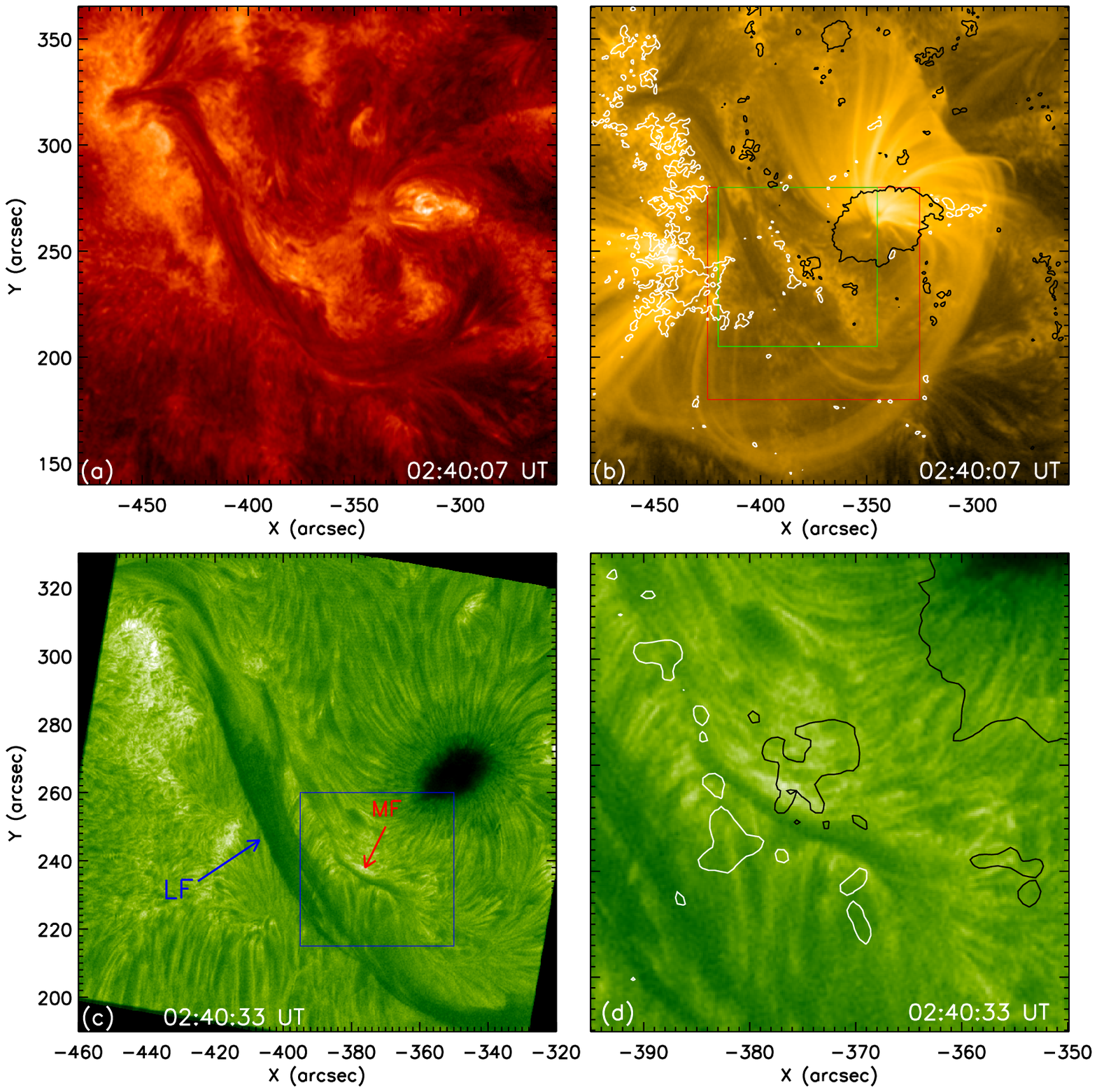}
\caption{Cotemporal AIA 304 \AA\ (panel ({\it a})), 171 \AA\ (panel ({\it b})) and NVST $H_{\alpha}$ line center (panel ({\it c})) images
showing the general appearance of the large filament ``LF" (as indicated by the blue arrow)
and the minifilament ``MF" (as denoted by the red arrow) at 02:40 UT on 2013 March 14.
Panel ({\it d}) shows a zoomed-in view corresponding to the blue box in panel ({\it c}).
Iso-Gauss contours of $\pm 200$ G are superposed by white and black curves in panels ({\it b}) and ({\it d}).
The red box shows the FOV of Figure 2({\it d})\sbond({\it f}) and ({\it h})\sbond({\it i}).
The green box indicates the FOV of Figure 3({\it a})\sbond({\it h}). An animation of panels ({\it c})\sbond({\it d}) is available.
The animation has a 11 s cadence, covering 02:30 UT to 05:30 UT.(
An animation of this figure is available.)}
\end{figure}

\begin{figure}
\epsscale{1.}
\plotone{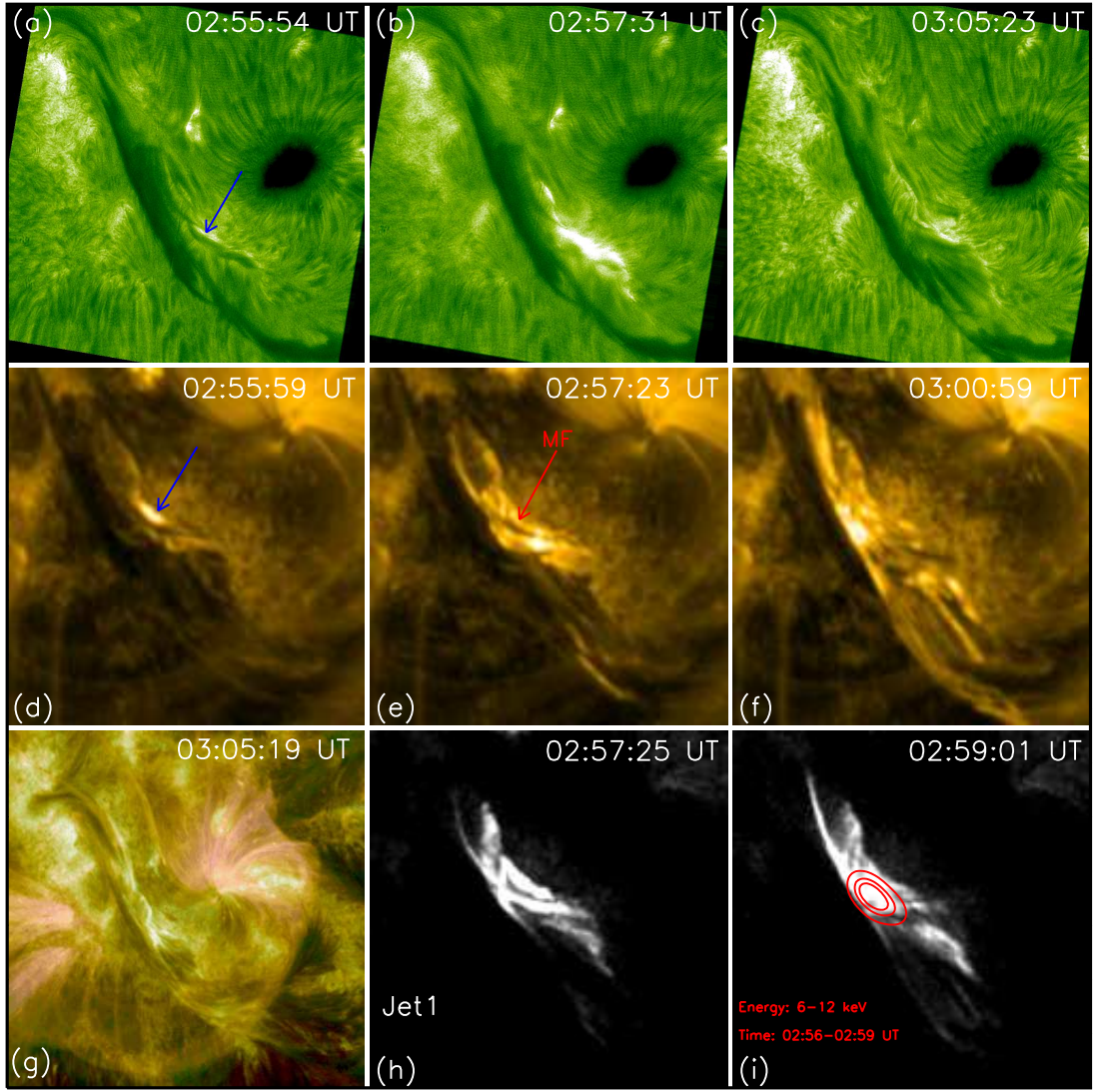}
\caption{Sequence of NVST $H_{\alpha}$ line center (panels ({\it a})\sbond({\it c})), AIA 171 \AA\ (panels ({\it d})\sbond({\it f})),
and 94 \AA\ (panels ({\it h})\sbond({\it i})) images exhibit MF eruption leading to the first two-sided loop jet.
(g) Composite image of the AIA 304 \AA\ (red) and 171 \AA\ (green) passbands displays the bi-directional jets moving along LF in opposite directions.
The blue arrows point to an initial brightening beneath MF as it just starts to rise, while the red arrow denotes the rising MF.
The red contours (80\%, 90\%, and 95\% of the maximum X-ray flux) in panel ({\it i}) represent \emph{RHESSI} 6-12 keV source.
Panels ({\it a})\sbond({\it c}) and panel ({\it g}) have the same FOV as Figure 1({\it c}) and Figure 1({\it a}), respectively.
An animation of panels ({\it d})\sbond({\it i}) is available. The animation has a 8 s in cadence,
covering 02:30 UT to 04:00 UT.
(An animation of this figure is available.)}
\end{figure}

\begin{figure}
\epsscale{1.}
\plotone{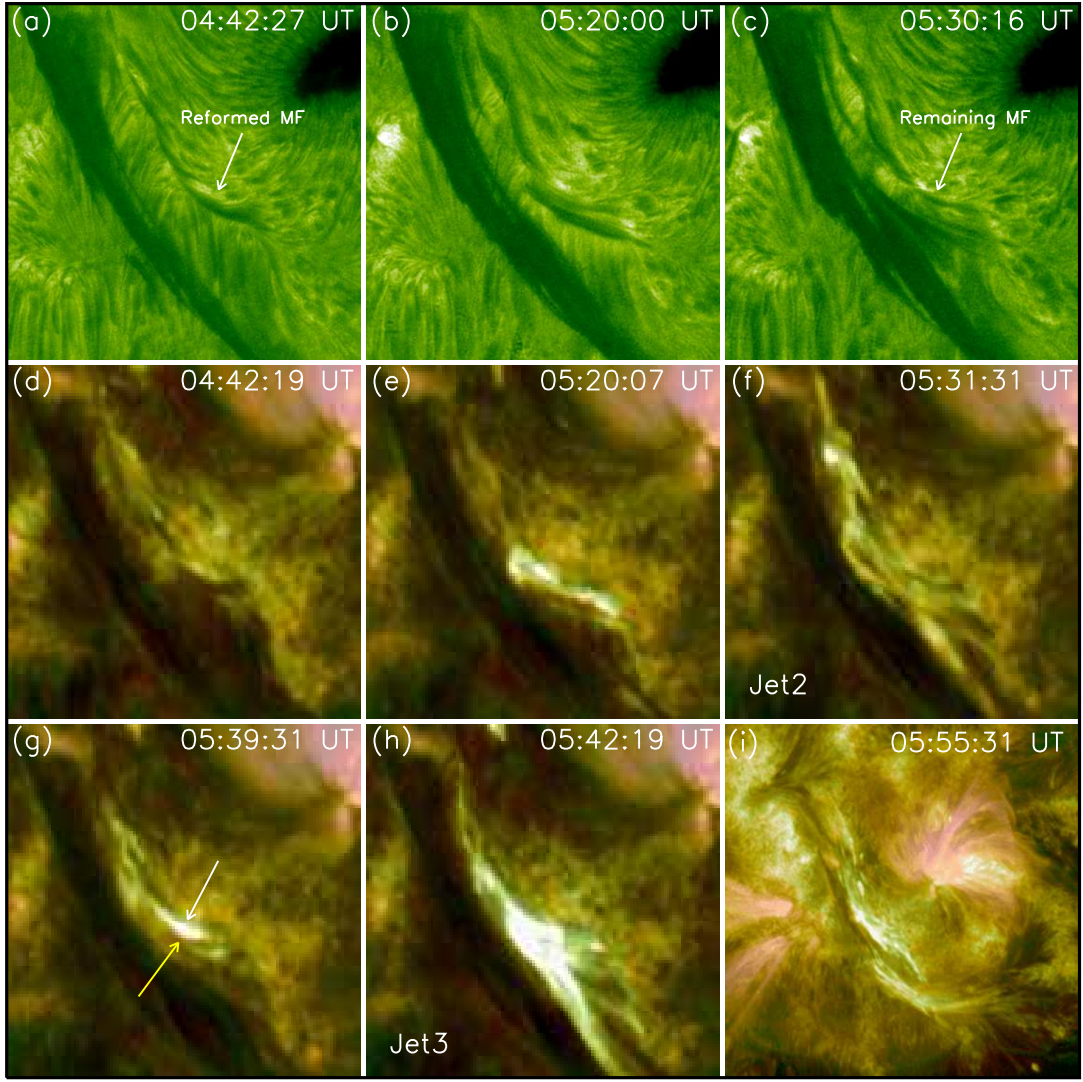}
\caption{Sequence of NVST $H_{\alpha}$ line center images (panels ({\it a})\sbond({\it c})) and composite images (panels ({\it d})\sbond({\it i}))
of the AIA 304 \AA\ (red) and 171 \AA\ (green) passbands display the partial and total eruptions of the reformed MF, resulting in the recurred two-sided loop jets.
The white arrow in panel ({\it a}) points to the reformed MF, in panel ({\it c}) points to the remaining MF after the partial eruption, and in panel ({\it g}) indicates
the brightening underneath the remaining MF. The yellow arrow in panel ({\it g}) denotes the rising of the remaining MF.
Panel ({\it i}) has the same FOV as Figure 1({\it a}).
An animation of panels ({\it d})\sbond({\it i}) is available. The animation has a 14 s in cadence,
covering 04:00 UT to 06:20 UT.
(An animation of this figure is available.)}
\end{figure}

\begin{figure}
\epsscale{1.1}
\plotone{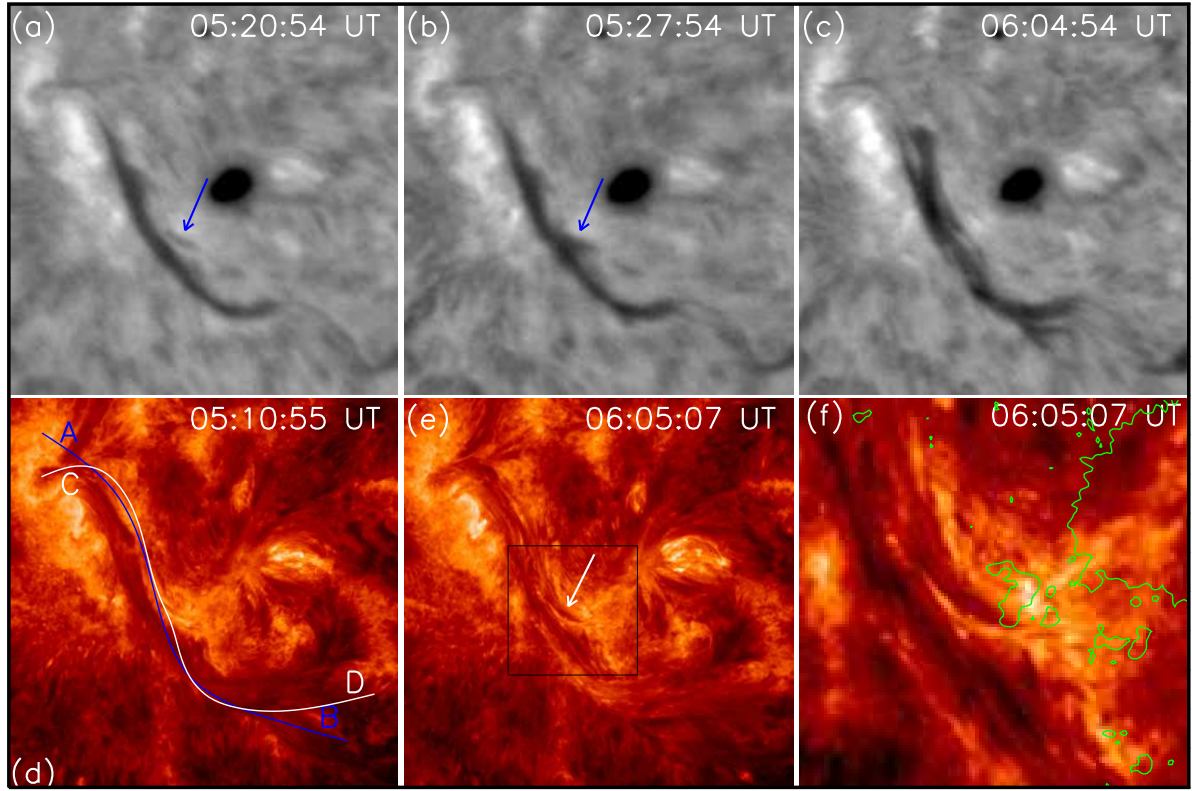}
\caption{GONG $H_{\alpha}$ images (panels ({\it a})\sbond({\it c})) present the eruption of the reformed MF.
The change of the magnetic field connectivity of LF after the eruption of the reformed MF is shown by the AIA 304 \AA\ images (panels ({\it d})\sbond({\it f})).
The blue arrows trace the rising of the reformed MF, while the white arrow indicates that
partial of LF's magnetic field connected to the negative ends of the reformed MF.
The blue curve ``AB" marks the slit position of the time slice shown in Figure 5({\it a})\sbond({\it b}),
while the white curve ``CD" marks the slit position of the time slice shown in Figure 5({\it c})\sbond({\it d}).
Iso-Gauss contours of -100 G are superposed by green curves on panel ({\it f}).
Panels ({\it a})\sbond({\it e}) have the same FOV as Figure 1({\it a}), and the black box shows the FOV of panel ({\it f}).}
\end{figure}

\begin{figure}
\epsscale{1.1}
\plotone{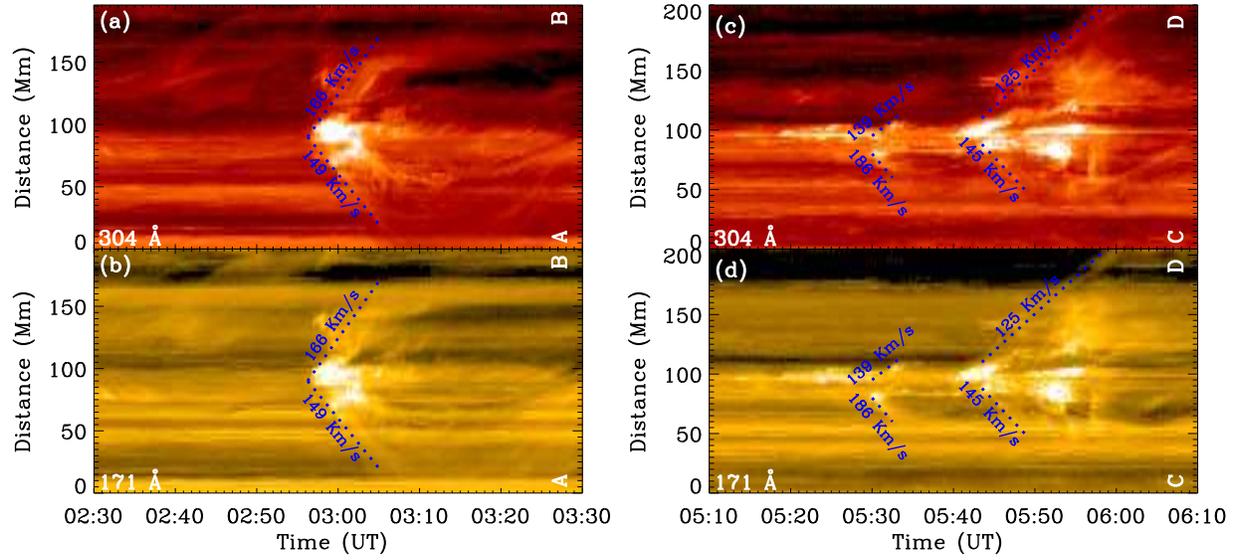}
\caption{Time-slice plots for the propagation of the two-sided loop jets along the blue curve
AB (({\it a})\sbond({\it b})) and the white curve CD (({\it c})\sbond({\it d})), respectively.
The declining dashed lines are linear fits to the paths of the two-sided loop jets.}
\end{figure}

\begin{figure}
\epsscale{1.1}
\plotone{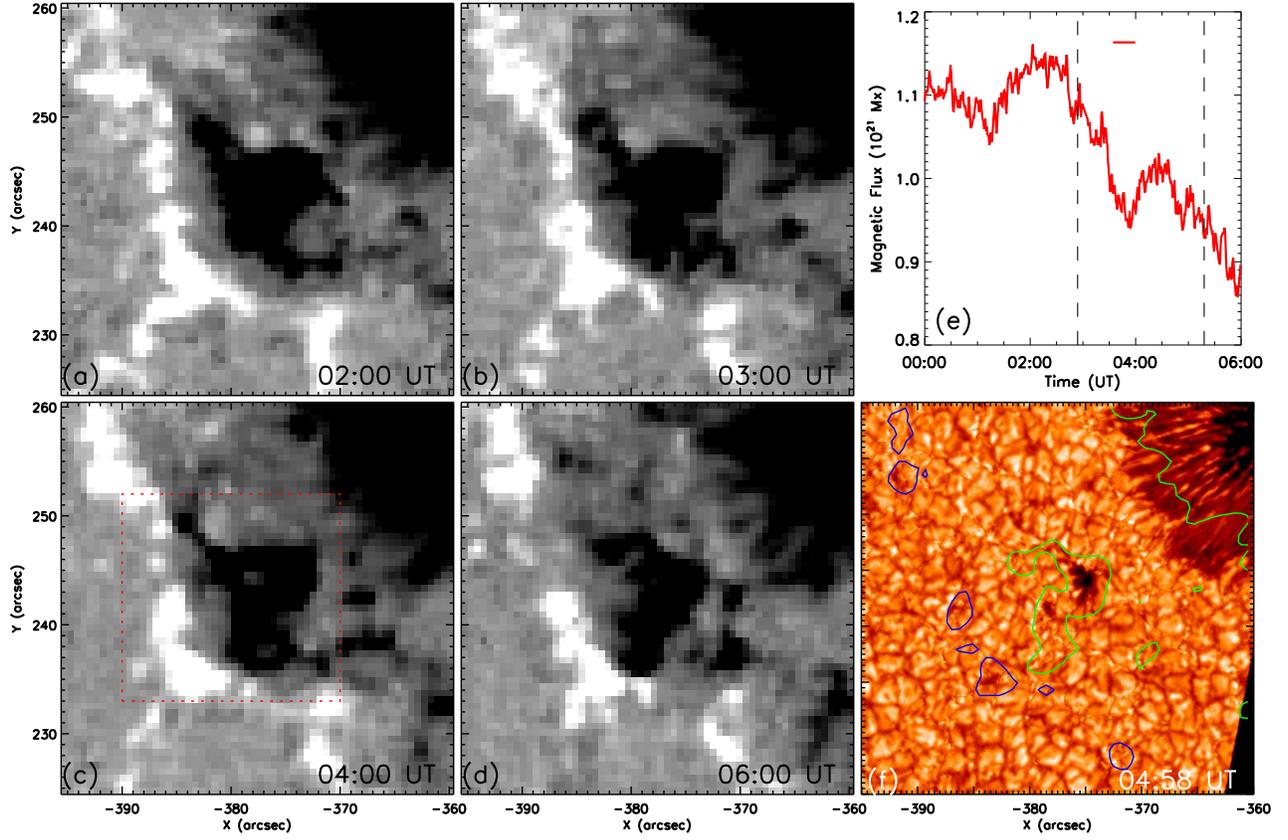}
\caption{\emph{SDO}/HMI vertical images (panels ({\it a})\sbond({\it d})) showing the cancellation of opposite polarities under MF.
The flux-time plot (panel ({\it e}) presents the temporal evolution of the negative magnetic flux, which is measured in a
region showed by the red dotted box in panel ({\it c}).
The vertical dashed lines indicate MF eruption onset times.
Iso-Gauss contours of $\pm 300$ G are superposed by blue and green curves on the NVST TiO image (panel ({\it f})).
}
\end{figure}
%\include{ref}

\end{document}